\documentclass[sigconf]{acmart} 

\AtBeginDocument{%
  \providecommand\BibTeX{{%
    \normalfont B\kern-0.5em{\scshape i\kern-0.25em b}\kern-0.8em\TeX}}}

\copyrightyear{2022} 
\acmYear{2022} 
\setcopyright{acmcopyright}\acmConference[VRCAI '22]{The 18th ACM SIGGRAPH International Conference on Virtual-Reality Continuum and its Applications in Industry}{December 27--29, 2022}{Guangzhou, China}
\acmBooktitle{The 18th ACM SIGGRAPH International Conference on Virtual-Reality Continuum and its Applications in Industry (VRCAI '22), December 27--29, 2022, Guangzhou, China}
\acmPrice{15.00}
\acmDOI{10.1145/3574131.3574459}
\acmISBN{979-8-4007-0031-6/22/12}

\usepackage{color, colortbl}
\begin{document}

\title{Exploring Text Selection in Augmented Reality Systems}

\author{Xinyi Liu}
\email{Xinyi.Liu1802@alumni.xjtlu.edu.cn}
\authornote{Both authors contributed equally to the paper.}
\affiliation{%
  \institution{Xi'an Jiaotong-Liverpool University}
  \city{Suzhou}
  \country{China}
}

\author{Xuanru Meng}
\email{ Xuanru.Meng18@alumni.xjtlu.edu.cn}
\authornotemark[1]
\affiliation{%
  \institution{Xi'an Jiaotong-Liverpool University}
  \city{Suzhou}
  \country{China}
}

\author{Becky Spittle}
\email{Becky.Spittle@bcu.ac.uk}
\affiliation{%
  \institution{Birmingham City University}
  \city{Birmingham}
  \country{UK}}

\author{Wenge Xu}
\email{Wenge.Xu@bcu.ac.uk}
\affiliation{%
  \institution{Birmingham City University}
  \city{Birmingham}
  \country{UK}
}

\author{BoYu Gao}
\email{bygao@jnu.edu.cn}
\affiliation{%
 \institution{Jinan University}
  \city{Guangzhou}
  \country{China}}

\author{Hai-Ning Liang}
\email{HaiNing.Liang@xjtlu.edu.cn}
\authornote{Corresponding author}
\affiliation{%
 \institution{Xi'an Jiaotong-Liverpool University}
  \city{Suzhou}
  \country{China}}

\renewcommand{\shortauthors}{Liu, Meng et al.}

\begin{abstract}
Text selection is a common and essential activity during text interaction in all interactive systems. As Augmented Reality (AR) head-mounted displays (HMDs) become more widespread, they will need to provide effective interaction techniques for text selection that ensure users can complete a range of text manipulation tasks (e.g., to highlight, copy, and paste text, send instant messages, and browse the web). As a relatively new platform, text selection in AR is largely unexplored and the suitability of interaction techniques supported by current AR HMDs for text selection tasks is unclear. This research aims to fill this gap and reports on an experiment with 12 participants, which compares the performance and usability (user experience and workload) of four possible techniques (Hand+Pinch, Hand+Dwell, Head+Pinch, and Head+Dwell). Our results suggest that Head+Dwell should be the default selection technique, as it is relatively fast, has the lowest error rate and workload, and has the highest-rated user experience and social acceptance.
\end{abstract}



\begin{teaserfigure}
  \includegraphics[width=\textwidth]{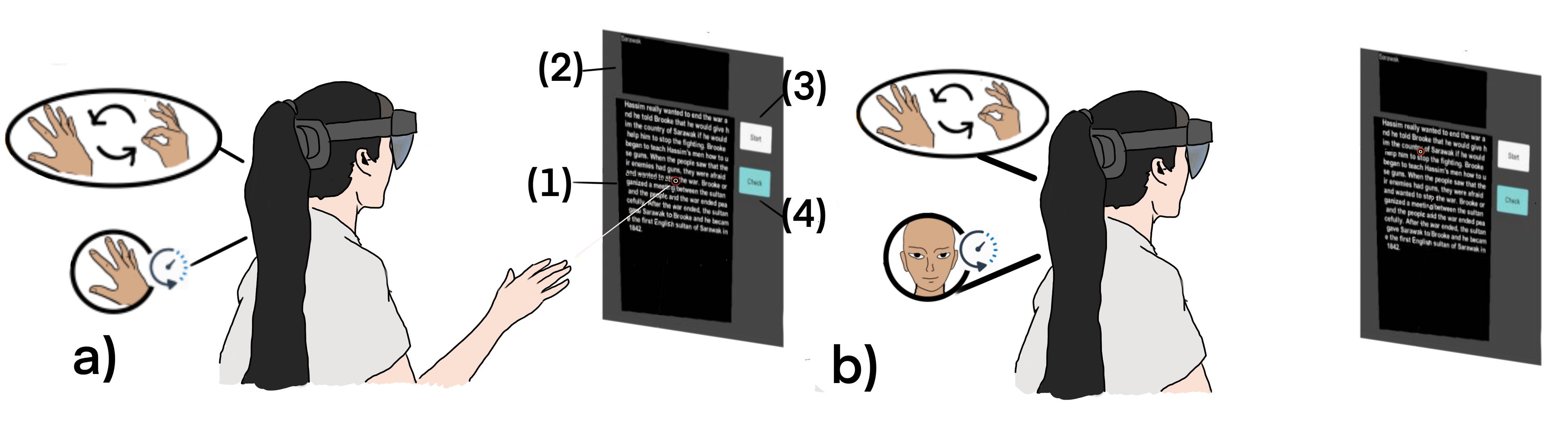}
  \caption{The four text selection techniques explored in this research: (a) \textit{Hand}-based: Hand+\textit{Pinch} and Hand+\textit{Dwell}, (b) \textit{Head}-based: Head+\textit{Pinch} and Head+\textit{Dwell}. The experimental interface used in this research: (1) an ‘interaction panel’, (2) an ‘instruction panel’, (3) the ‘start’ button, and (4) the ‘check’ button. }
  \Description{The four text selection techniques explored in this research: (a) \textit{Hand}-based: Hand+\textit{Pinch} and Hand+\textit{Dwell}, (b) \textit{Head}-based: Head+\textit{Pinch} and Head+\textit{Dwell}. Experimental interface setup: (1) an ‘interaction panel’, (2) an ‘instruction panel’, (3) the ‘start’ button, and (4) the ‘check’ button.}
  \label{fig:teaser}
\end{teaserfigure}

\maketitle

\section{Introduction}
Augmented reality (AR) head-mounted displays (HMDs) represent the next generation of mobile computing devices that can transform how people engage with digital information. Currently, AR users can read academic papers, make personal notes\footnote{https://www.microsoft.com/en-us/p/airnotes/9phsgp1kt88c?activetab=pivot:overviewtab}, have live meetings with colleagues\footnote{https://docs.microsoft.com/en-us/dynamics365/mixed-reality/remote-assist/overview-hololens}), and send instant messages. There has been some prior work exploring the use of AR to support tasks surrounding text manipulation \cite{8943748}. However, fundamental to any text manipulation task, including highlighting, modifying, and creating documents, is text selection, which is still largely unexplored on AR HMDs.  

This work investigates a basic but essential and common task related to text manipulation, i.e., text selection. Text selection is the first stage involved in text editing tasks \cite{8943748}, and is the process of selecting text fragments that need further manipulation (e.g., copying, cutting, or deleting) or making notes on a document. As a foundational activity, text selection can be used in a wide range of scenarios, such as (1) copying and pasting text between different applications (e.g., instant messaging apps), (2) highlighting text fragments for later reference while reading articles, and (3) editing text in text boxes and fields in online forms. Darbar et al. \cite{darbar2021exploring} have explored using an external handheld device (a smartphone) as an input medium for text selection, which has several limitations: (1) the current positional tracking of the smartphone is not practical (because of this, Darbar et al. had to use external optical-based tracking devices), and (2) it is dependent on having a handheld device at all times, which is not always convenient and practical. 

To overcome these limitations, this paper explores existing interaction techniques (hands-free or handheld-device-free---having no need for users to hold a device in their hands; see examples provided in  \autoref{fig:teaser}) that can be captured by the sensors that are built into commercial AR HMDs. Specifically, we focus on the combination of pointing methods and selection mechanisms \cite{8943748}, where the pointing method is used for identifying a text fragment to be selected, and the selection mechanism is used to initiate and finalize a selection~\cite{meng_handsfree_2022,xu_withhands_2022}. Although there have been studies on interaction techniques for tasks like text entry in VR \cite{8943748} and object acquisition in AR \cite{9284722}, these results cannot be applied directly to text selection. This is because (1) text selection tasks require users to hold the selection mechanism from the start of selection to the end, a process that requires effective precision, control, and focus over a period of time; (2) the density of text information is usually high because letters and sentences tend to be close to one another; and (3) given the limited field of view of AR HMDs, the letters are displayed in an even smaller area, which makes selections more difficult. 

We conducted a user study with twelve participants to evaluate the performance and usability of four interaction techniques (\textit{Hand+Dwell}, \textit{Hand+Pinch}, \textit{Head+Dwell}, \textit{Head+Pinch}) for AR HMDs (see \autoref{fig:teaser} for an illustration of the techniques). The two most prominent pointing methods used for AR HMDs are explored (\textit{head}-based and \textit{hand}-based), as well as two feasible selection mechanisms (Dwell and Pinch), which were chosen based on findings from prior work and results from our pilot tests. Our results suggest that \textit{Head+Dwell} is the leading candidate for text selection with AR HMDs since it is relatively fast, has the lowest error rate and workload rating, has the highest user experience rating, and is socially acceptable. 

The main contribution of this work is a foundational empirical study of four possible techniques for text selection on AR HMDs, regarding their performance, user experience, and workload. Our work provides insights and suggestions for AR HMDs to support text manipulation tasks for such devices.

\section{Related Work}

\subsection{Text Selection for AR/VR HMDs}
There has been some recent work on text selection for VR HMDs \cite{meng_handsfree_2022,xu_withhands_2022}. However, text selection in AR is more problematic and complex and differs from VR in several areas: (1) layer interference \cite{kruijff_perceptual_2010}; (2) low visibility, due to screen brightness and contrast, or color and texture patterns that can interfere with the captured environment \cite{kruijff_perceptual_2010}; and (3) layout foreground-background issues \cite{8943748}. In addition, AR HMDs require smooth, efficient tracking that is capable of maintaining robustness under diverse environmental conditions, such as changing light levels \cite{peddie2017augmented}. These factors affect text readability and visibility and can introduce scene distortion, hence making it more difficult to complete text selections \cite{kruijff_perceptual_2010}. 

There have been some studies exploring the use of other external devices for text selection with AR HMDs. For instance, EYEditor \cite{ghosh2020eyeditor} uses a ring mouse for cursor navigation and text selection, where a button is used for placing the cursor before and after a text fragment to be selected, while the selection is made via a touchpad. Researchers have also investigated using a smartphone for input on AR HMDs. For example, Lee et al. \cite{lee_one-thumb_2020} have used a force-sensitive smartphone as their input device, and users exert a force on a thumb-sized circular button to select a desired text fragment. Similarly, Darbar et al. \cite{darbar2021exploring} have explored several methods (continuous touch, discrete touch, spatial movement, and ray casting) that rely on a smartphone for text selection with AR HMDs and found that continuous touch is more efficient than the other three methods. However, just like controller-based input, these proposed external input devices are not that convenient, need extra external hardware, and are often not efficient and usable \cite{ghosh2020eyeditor}. In general, because there has been limited research, there is still a lack of understanding of what techniques are supportive regarding text selection tasks with AR HMDs, especially those that do not require any additional input device but rely solely on the hardware features available on the HMDs themselves.

\subsection{Handheld-device-free Pointing-based Interaction with AR HMDs} \label{interactionDef}

\subsubsection{Head-based}
Head-based pointing has been widely adopted as a standard method for pointing at virtual objects in HMDs since it is accurate, comfortable, and convenient, especially when handheld pointing devices are not available \cite{kyto2018pinpointing} but lacks an intrinsic mechanism for users to confirm an identified target \cite{ESTEVES2020102414}. One selection method that has been widely coupled with Head-based pointing is the Dwell technique. It has been recommended as the first option in both handheld-device free and hands-free scenarios for text selection for VR HMDs \cite{xu_withhands_2022}. Although it is less error-prone \cite{xu_withhands_2022}, it is comparably slow and the predefined dwell time might be perceived to be stressful for users during interaction, as they often feel that they are continually being pushed towards making the next selection quickly.

Performing bare hand gestures, such as a Pinch (i.e., closing the thumb and index finger together), is another handheld-device-free method for indicating a selection and can be used with Head-based pointing during interaction \cite{xu_withhands_2022}. Prior work suggests that using a Pinch gesture is as effective as clicking a button on a handheld controller and is faster than Dwell for object acquisition tasks with eye-gaze pointing \cite{mutasim2021pinch}. The negative aspect of Pinch is that users tend to leave their hand posture to a "comfort grip", which can unintentionally activate selections, leading to high errors, whereas Dwell has no such issues and error types \cite{pfeuffer2017gaze+}. In addition, using a hand gesture to indicate a selection has been found to cause higher fatigue than Dwell \cite{yan2020headcross}.

\subsubsection{Hand-based}
Hand-based pointing is now possible on several AR and VR HMDs and is perceived to be a natural and intuitive interaction method. Users can use hand-based pointing to point at distant objects and then perform a hand gesture to indicate a selection (typically a Pinch gesture). This method has a short learning curve, but users tend to leave their hand posture in a "comfort grip" \cite{pfeuffer2017gaze+}, as described previously, which can lead to more errors. In this case, users can use Dwell to prevent the types of errors caused by the "comfort grip" tendency. In addition, it has been observed that due to the relatively small hand tracking field of view and users' unfamiliarity with new technology, novice users may inadvertently move their hands outside the hand tracking area captured by an AR HMD's tracking sensors, which can lead to mistakes during interaction \cite{arboundary}.

\subsubsection{Eye Gaze-based}
Eye tracking is becoming a standard feature of AR HMDs (e.g., HoloLens 2 and Magic Leap series). It is a fast and effortless contextual input and it is most powerful when combined with other selection methods such as Voice or Pinch to confirm the user's intent~\footnote{https://docs.microsoft.com/en-us/windows/mixed-reality/design/eye-gaze-interaction}. However, we did not consider gaze because of its inaccuracy in selecting small targets (text letters are smaller than the recommended 2$^{\circ}$ in reading angle)\footnotemark[3] in dense areas (as texts are also usually placed very close to each other). We tested the accuracy of the HoloLens 2's eye tracking sensors for text selection with five pilot users and confirmed that, as described on the official website, it was not that stable for fixations when reading texts within the small AR display.

\subsubsection{Other Handheld-device-free Selection Methods}

\textit{Eye blinking} has been recently used for confirming selections \cite{lu2020exploration, lu202iItext} and has been found to be a viable solution for confirming a text selection for VR HMDs \cite{meng_handsfree_2022}. To determine whether it is also feasible for AR HMDs, we implemented eye blinks \cite{lu2020exploration, lu202iItext} and asked our five pilot users to test the technique. Results revealed that it is hard for users to stop their eyes from blinking while trying to look at the AR screen carefully, especially for long text fragments. Users could adequately indicate the starting selection point, but arriving at the end selection point was challenging, as they would often blink their eyes unconsciously prior to reaching the endpoint, thereby causing a large number of false positive activations. Given these results, we did not consider eye blinks further.

\textit{Voice} activation has been widely used as an input modality. Nevertheless, we did not include it because: (1) speech interaction is not very robust as the system may struggle to adapt to the range of inconsistencies present in natural language (which includes people's accents, dialects, and other pronunciation patterns) \cite{spittle_2022_a}; (2) the recognition can be affected by background noise when it is used in a public venue; and (3) using voice input may affect nearby users in the work environment (e.g., co-workers) and users found it to be not socially acceptable as a text confirming method \cite{meng_handsfree_2022}; our pilot study also confirmed it.

\textit{Neck Motions} such as forward and backward movements along the depth dimension (or z dimension from the users' perspective) have been explored for object/information activation tasks \cite{lu2020exploration}. However, we did not include this method as prior work suggests that (1) these types of motions are not precise enough when dealing with text content, (2) users dislike them as they found them tiring and uncomfortable to perform, and (3) users found it is not socially acceptable to perform in public \cite{meng_handsfree_2022}. Our pilot study with 5 users confirmed these findings.

\subsection{Inclusion Criteria for the Techniques} \label{Section:UsabilityCriteria}
We identified four usability criteria for the methods that can be used for confirming a selection (that can work well with both Head-based and Hand-based pointing). 

\begin{itemize}
    \item \textit{\textbf{C1: Simple, easy, and efficient to use}}. Text selection is a two-step process, where selection confirmation needs to be well coordinated with the pointer movement. With users controlling the pointer with their heads or hands while paying attention to the highlighted text, selection should be simple, easy to use, and fast/efficient, so that it is possible to perform the two tasks almost simultaneously.

    \item \textit{\textbf{C2: Minimal error rate and workload}}. Both Head-based and Hand-based interaction should be accurate, comfortable, and convenient to perform. As such, the ideal situation is that the selection mechanism minimizes any additional workload during text selection and has some degree of precision while allowing fast interaction and minimizing false activations, as recovering from errors during text selection can be difficult and frustrating.

     \item \textit{\textbf{C3: Accessible and mobility enabling}}. AR HMDs are projected to become ubiquitous smart devices that people will use to interact on the go. Consequently, they should be as portable as possible. Even when they are not used outdoors, people often employ them for mobile applications and scenarios \cite{spittle_2022_a}. As such, given that Head-based and Hand-based pointing can now be achieved without using external tracking devices, the selection mechanism should also be handheld-device-free and support users' mobility. 

     \item \textit{\textbf{C4: Socially acceptable}}. AR HMDs are meant to be highly portable and enable interaction with multiple users in one physical space. Therefore, as prior research has shown \cite{Fouad.2018.PerformerSocial, Pandey.2021.Speech, Vergari.2021.SocialEnv}, the social acceptability of interactions used for AR HMDs can be an important factor in determining their widespread adoption and usability. As such, the interactions should also be socially acceptable and transferable to a range of situations and use cases.

\end{itemize}

\begin{table}
\centering
  \caption{Ratings for each criterion regarding the five possible selection mechanisms identified from the literature (0-3 ticks indicate ratings from worst to the best). (C1) Simple, easy, and efficient to use; (C2) Minimal error rate and workload; (C3) Accessible and mobility enabling; and (C4) Socially acceptable.}
  \label{tab:freq}
  \begin{tabular}{p{1.5cm}p{1.2cm}p{1.2cm}p{1.2cm}p{1.2cm}}
    \toprule
    Mechanism& C1 & C2 & C3 & C4\\
    \midrule
    Dwell & \checkmark \checkmark \checkmark & \checkmark \checkmark \checkmark & \checkmark \checkmark \checkmark & \checkmark \checkmark \checkmark \\
    Pinch & \checkmark \checkmark \checkmark & \checkmark \checkmark & \checkmark \checkmark \checkmark & \checkmark \checkmark \\
    Eye blinks & \checkmark \checkmark \checkmark & - & \checkmark \checkmark \checkmark & \checkmark \checkmark \checkmark \\
    Voice & \checkmark \checkmark & \checkmark \checkmark & \checkmark \checkmark \checkmark & - \\
    Neck & - & \checkmark & \checkmark \checkmark \checkmark & - \\

\bottomrule
\end{tabular}
\label{table:CriteriaPre}
\end{table}

The criteria and the decision as to whether they should be chosen were based on the literature review and pilot tests (see above sections). \autoref{table:CriteriaPre} shows the ratings of the five possible selection methods found in the literature against the four usability criteria. The ratings are based also on the results of prior work together with results from our tests with 5 pilot users. To qualify as a suitable candidate, a mechanism should have at least one tick for each of the four criteria. While they meet C3 (mobility) well, only Dwell and Pinch passed the overall evaluation. Both Voice and Neck did poorly in C4 (social acceptability) and because of this, as well as other weaknesses, they were not considered in our experiment. Eye blinks were considered to be a good candidate based on findings from prior work, especially in text entry. However, we found that they were not suitable for text selection due to poor ratings in C2 (error rate and workload).

\section{Experiment}
\subsection{AR Testbed Environment}
The testbed environment was developed using Unity3D (v2019.4.30). \autoref{fig:teaser} shows the environment used in this experiment. An `interaction panel' is located in the center, from where the participant could use an interaction technique to select a text fragment. Above the `interaction panel', there is an `instruction panel' showing the target text that needs to be selected. There are two buttons on the right-hand side, where clicking the `start' button would start the timer, and clicking the `check' button would check if the selection is correct and record the time. Participants would only move on to the next trial when a correct selection was made.

The parameters of the environment were defined according to recommendations found in AR and VR literature and were determined through a pilot study with five pilot testers. The length of the material was kept between 35 and 40 characters per line (including symbols and spaces) \cite{wei2020reading}. A dark color was used as the background for the panels \cite{10.1145/3357251.3357584}, and contrasting colored text was overlaid using the Arial typeface with 1.8$^{\circ}$ angular size to maximize legibility \cite{dingler2018vr}. The main interaction panel was located 2 meters in front of the user. The pilot users confirmed that they were able to read the text clearly without any difficulties. Also, the interface elements were clear and easy to understand, and thus we ensured there were no issues with environmental lighting and contrast conditions.

\subsection{Evaluated Interaction Techniques}
All of the following techniques were developed using the default input functions provided by the MRTK Unity Plugin (v2.7).

\subsubsection{Head-based Pointing}
A cursor was attached to the end of an invisible ray, which extended from the HMD towards the viewing direction (see \autoref{fig:teaser}.b).

\textit{Head-based pointing + Dwell (Head+D)}. The user started the selection of the desired text fragment by dwelling (maintaining cursor position for one second) directly before the first letter and completed the selection by dwelling (one second) again at the end of the text fragment. The dwell duration was determined through our pilot study, which tested a range of thresholds from 400ms to 1s.

\textit{Head-based pointing + Pinch (Head+P)}. The user started the selection by positioning the head cursor at the beginning of the text fragment and performing the ``Pinch” gesture. They then held the gesture until the head pointer moved to the end of the target text fragment, and released the Pinch gesture to end the selection.

\subsubsection{Hand-based Pointing}
\autoref{fig:teaser}.a shows the two hand-based techniques implemented in our study. A cursor was attached to the end of a ray rendered from the palm. We controlled the lighting of the physical environment, based on the recommendations of the HoloLens 2 official website~\footnote{https://docs.microsoft.com/en-us/hololens/hololens-environment-considerations}, to ensure the hand tracking was reliable and consistent throughout the study.

\textit{Hand-based pointing + Dwell (Hand+D)}. Hand+D was analogous to Head+D, but employed Hand-based pointing with a Dwell time.

\textit{Hand-based pointing + Pinch (Hand+P)}. Hand+P was analogous to Head+P, but used Hand-based pointing in combination with Pinch. 

To deliver a good interaction experience, we retained the default support and feedback provided by the MRTK plugin. Specifically, we employed the following implementation: (1) the hand ray visualization was enabled whereas the head ray was invisible; (2) the end of the ray was akin to a donut-shaped cursor to indicate the location where the ray intersected with a target object. The cursor would change from being hollow to being filled to indicate a selection had been made, with both Dwell and Pinch. In addition, we implemented the following support and feedback for each selection: (1) the selected text was highlighted in yellow; (2) the dwell time was displayed as a loading circle; and (3) voice prompts were provided ``Your selection was incorrect, please re-select" if the text was selected incorrectly.

\subsection{Design and Task}
\label{sec:design/task}
The experiment followed a within-subjects design with Interaction Techniques (Head+D, Head+P, Hand+D, Hand+P) as the independent variable. For each condition, participants completed a series of selection tasks, where they were required to make text selections for different Sentence Lengths: (1) short (1 word) and (2) long (5-6 lines) fragments. Overall, participants completed 8 training trials (4 short, 4 long text fragments) and 20 experimental trials (10 short, 10 long text fragments). Text fragments were randomly selected from a corpus of standardized English reading assessments \cite{quinn2007asian}. To avoid learning effects, each text fragment appeared only once in a given condition. The order of the Sentence Length was random, and the order of the Interaction Technique was counterbalanced across participants. A total of 960 experimental trials were collected (12 participants × 4 interaction techniques × 20 selections). 


\subsection{Participants and Apparatus}
A total of 12 university students (6 males; 6 females), aged between 19 and 23 years were recruited for this experiment (M=21.4, SD=1.00). They were all right-handed, had normal or corrected-to-normal vision, and had no difficulties with arm and head movements (which we checked prior to the experiment). Ten participants had previous experience with VR HMDs, and four of them were regular users (daily or weekly). Nine participants had previous experience with AR HMDs (they had used them in the last 3-6 months), but none were frequent users. During the experiment, participants were seated in an office chair and were allowed to utilize the armrests. The HoloLens 2 HMD was used to conduct our study.

\subsubsection{Measurements}
To assess the text selection performance and experience, we collected the following measurements:
\begin{itemize}
\item \textit{Objective}. (a) \textit{Task completion time}: Time from when the participant clicked the start button to the time the participant clicked the check button (when the correct selection was made); (b) \textit{Error rate}: the number of incorrect selections (recorded by the Check button) compared to the total number of selections.

\item \textit{Subjective}.  (a) \textit{Workload}: NASA-TLX questionnaire \cite{hart1988development}, and (b) \textit{User Experience}: User Experience Questionnaire (UEQ) \cite{laugwitz2008construction}. \textit{Qualitative feedback}, regarding the advantages and disadvantages of each interaction method, was also captured, together with \textit{user rankings} of the text selection techniques.
\end{itemize}

\subsection{Procedure}
Prior to the experiment, participants were informed of the purpose of the research and filled out a demographics questionnaire to collect personal anonymized information (i.e., age, experience with AR/VR, etc.). Before the start of each condition, the corresponding text selection technique was explained to the participants, and they were required to undergo training where they completed 8 trials for each technique (4 short texts and 4 long texts). For the formal trials, participants needed to complete 20 text selections for each condition (as described in section \ref{sec:design/task}). Participants were allowed to rest for as long as they wanted between conditions. At the end of each condition, participants completed the NASA-TLX and UEQ. At the end of the experiment, they participated in a structured interview that collected their preferences and rankings of the four techniques, plus any other feedback they had.

\subsection{Results}
We used Shapiro-Wilks tests and Q-Q plots to check if the data was normally distributed. 

\paragraph{Performance analysis} For normally distributed data, two-way repeated measures ANOVAs were used with \textit{Interaction Techniques} (Head+D, Head+P, Hand+D, Hand+P) and \textit{Sentence Lengths} (short, long) as within-subjects variables. For non-normally distributed data, we processed the data through an Aligned Rank Transform (ART) \cite{Jacob_ART} before using repeated-measures ANOVAs with the transformed data. \textit{Experience analysis}. For normally distributed data, we used one-way repeated measures ANOVAs with Interaction Techniques as the within-subjects variable. Otherwise, we first processed the data through ART and then used repeated measures ANOVAs with the transformed data.  

For both analyses, we used Bonferroni correction for pairwise comparisons and Greenhouse-Geisser adjustment for degrees of freedom where the assumption of sphericity was violated. All tests were conducted with two-tailed p-values.

\subsubsection{Performance}

\paragraph{Task completion time}
There was a statistically significant difference in the completion time for each Sentence Length ($F_{1,77}=286.251, p<.001$). Post-hoc analysis with Bonferroni correction suggested that, for all the techniques (all $p<.0001$), the completion time for long-text tasks was longer than for short-text tasks. There was no significant main effect of Interaction Techniques ($F_{3,77}=1.69, p=.175$) on task completion time, or an effect of Interaction Techniques $\times$ Sentence Lengths ($F_{3,77}=0.79, p=.504$).

\paragraph{Error rate}
A two-way repeated measures ANOVA yielded a significant main effect of Interaction Techniques ($F_{3,77}=4.49, p<.01$) on error rate. Post-hoc analysis with Bonferroni correction suggested that Head+D outperformed Hand+D ($p<0.05$). We could not find a significant main effect of Sentence Lengths ($F_{1,77}=0.68, p=.412$) or any effect of Interaction Techniques $\times$ Sentence Lengths ($F_{3,77}=1.10, p=.352$). Details of error rate results can be found in \autoref{table:performance}. 

\begin{table*}
  \caption{Performance data for each Interaction Technique for two Sentence Lengths, mean (SD). The top 3 techniques of each condition are presented with Roman numerals (I: yellow; II: purple; and III: green).}
  \label{tab:freq}
  \begin{tabular}{p{3cm} p{2.2cm} p{2.1cm}p{2.5cm}p{1.98cm}p{2.5cm}p{2.33cm}p{2.33cm}}
    \toprule
    Performance Metrics&Sentence Length&Head+D&Head+P&Hand+D&Hand+P\\
    \midrule
    Task completion time   & Short& \cellcolor{green!30}III: 4.56 (1.98) & \cellcolor{yellow}I: 3.86 (1.57) & 5.06 (2.50) & \cellcolor{blue!30}II: 4.47 (1.82) \\
    & Long & \cellcolor{yellow}I: 7.59 (2.78) & \cellcolor{blue!30}II: 7.93 (2.72) & 8.43 (3.33) & \cellcolor{green!30}III: 8.23 (2.70)\\
    
    Error rate & Short & \cellcolor{yellow}I: 6.34\%  (7.41\%) & \cellcolor{green!30}III: 16.96\% (10.68\%) & 23.79\% (15.32\%) & \cellcolor{blue!30}II: 11.67\% (10.58\%)\\
    &Long&\cellcolor{yellow}I: 10.00\% (6.23\%) & \cellcolor{green!30}III: 14.06\% (8.51\%) & 15.6\% (13.08\%) & \cellcolor{blue!30}II: 11.40\% (9.18\%)\\

\bottomrule
\end{tabular}
\label{table:performance}
\end{table*}

\subsubsection{Experience}

\paragraph{UEQ}
The average UEQ scores for each Interaction Technique are Hand+D (M=-0.06, SD=0.31), Hand+P (M=-0.01, SD=0.14), Head+D (M=0.27, SD=0.33), Head+P (M=0.10, SD=0.26). ANOVA tests revealed a significant difference between Interaction Techniques on the average UEQ scores ($F_{2.638,187.270} = 14.818, p < .001$), with post-hoc pairwise comparison tests suggesting that: (1) Head+D outperformed Hand+D ($p < .001$) and Hand+P ($p < .01$); (2) Head+P performed better than Hand+D ($p < .001$); and (3) Hand+P outperformed Hand+D ($p < .05$). 

\begin{figure}[t]
    \centering
    \includegraphics[width=1\linewidth]{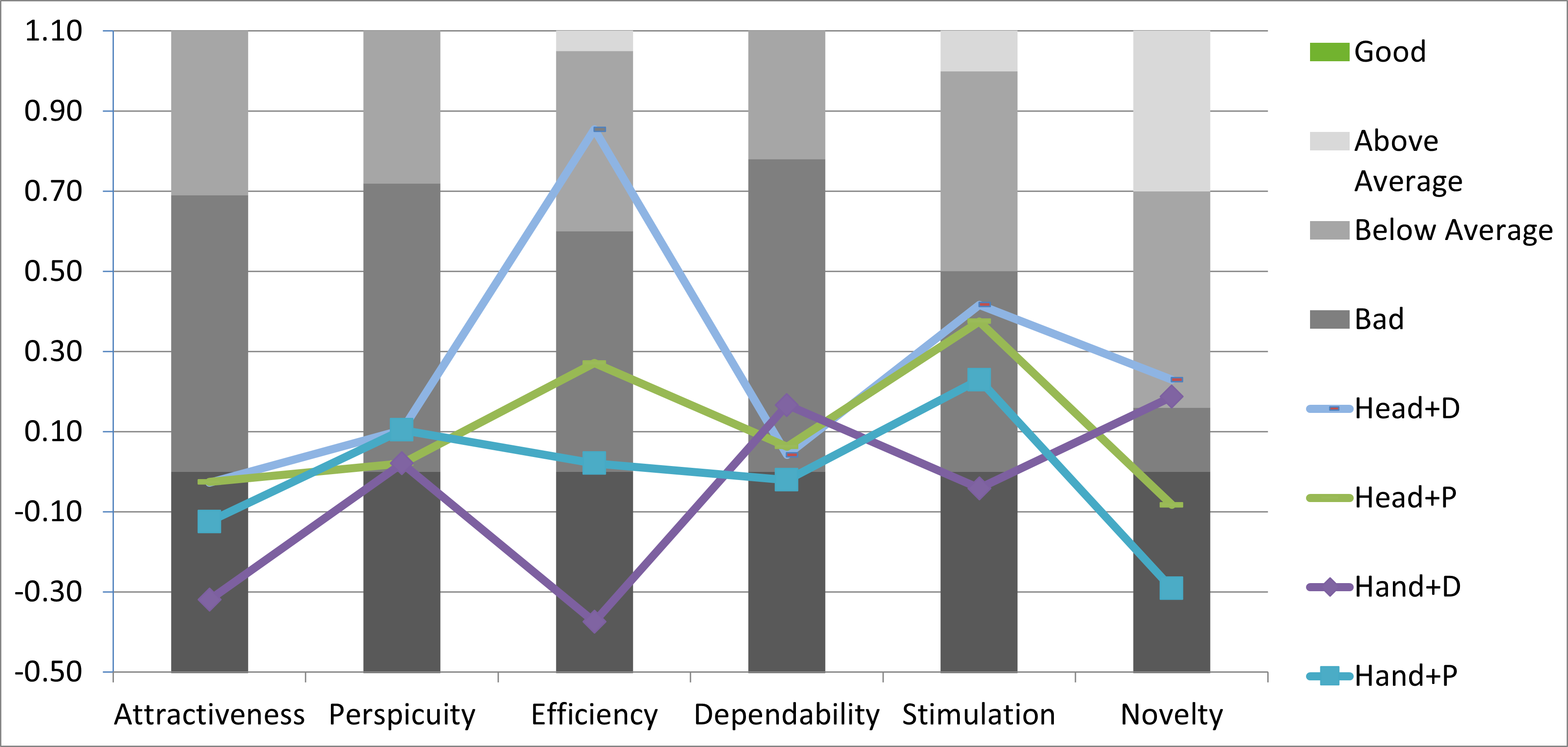}
    \caption{UEQ subscale ratings of the evaluated techniques with respect to the relative benchmark values.}
    \label{fig_UEQS}
\end{figure}

Regarding the UEQ subscales, there was a significant effect of Interaction Techniques on Efficiency ($F_{3,33} = 9.007, p < .005$). Post-hoc tests indicated that Head+D had a higher Efficiency than Hand+D ($p < .005$) and Hand+P ($p < .005$). ANOVA tests also suggested a significant effect of Interaction Techniques on Stimulation ($F_{3,33} = 3.056, p < 0.05$). However, we could not find any significant difference between Interaction Techniques based on the results from the post-hoc pairwise comparisons. No other significant effect were found on Attractiveness ($F_{3,33} = 2.154, p = 0.112$), Perspicuity ($F_{3,33} = 0.90, p = 0.965$), Dependability ($F_{3,33} = 0.391, p = 0.760$), and Novelty ($F_{3,33} = 2.855, p = 0.052$). Details of each UEQ subscale score can be found in \autoref{fig_UEQS}. 


\paragraph{Workload} ANOVA tests revealed a significant effect of Interaction Techniques ($F_{3,33} = 7.098, p < .005$) on the overall workload. Pairwise post-hoc comparisons suggested that Head+D required less workload than Hand+D ($p < .01$). The overall workload for each Interaction Technique is: Hand+D (M=48.42, SD=17.68), Hand+P (M=37.61, SD=13.76), Head+D (M=24.17, SD=13.94), Head+P (M=34.69, SD=23.45).  

\begin{figure}[t]
    \centering
    \includegraphics[width=1\linewidth]{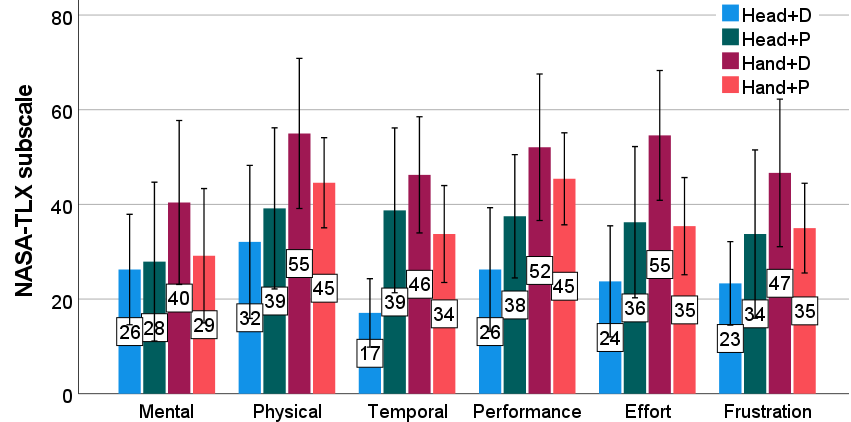}
    \caption{The mean responses for the 6 subscales of the NASA TLX questionnaire. Error bars indicate a 95\% confidence interval.}
    \label{fig_workloadS}
\end{figure}

For NASA-TLX workload subscales, we observed a significant effect of Interaction Techniques on Temporal Demand ($F_{3,33}=5.717, p <.01$) and Effort ($F_{3,33}=9.072, p<.001$). Post-hoc pairwise comparisons revealed that Head+D had lower Temporal Demand and Effort than Hand+D (both $p<.01$). ANOVA tests also indicated that there were significant effects of Interaction Techniques on Performance ($F_{3,33}=5.298, p <.01$) and Frustration ($F_{3,33}=3.303, p<.05$). Post-hoc tests showed that Head+D had a significantly better performance than Hand+P ($p<.05$). However, we could not find any significant difference between techniques on Frustration. In addition, ANOVA tests yielded no significant effect of Interaction Techniques on Mental ($F_{3,33} = 1.492, p = 0.235$) and Physical Demand ($F_{3,33} = 2.787, p = 0.056$). \autoref{fig_workloadS} shows the detailed results of the NASA-TLX subscales.

\paragraph{Ranking} Head+D was the most popular technique, with ten participants ranking it first and one ranking it second. This was followed by Head+P; one participant ranked it first and seven ranked it second. The rankings of the Hand-based techniques were generally worse, with the majority of participants ranking them third (four voted for Hand+D and five for Hand+P) or fourth (five voted for Hand+D and another five for Hand+P).

\subsubsection{Qualitative Feedback} 
Feedback for Head-based techniques was mostly positive. For instance, Head+D was perceived to be ``\textit{easy and simple}" (N=7), ``\textit{fast}" (N=5), and ``\textit{stable}" (N=2). For Head+P, participants considered it as ``\textit{novel}" (N=7). However, there was also some negative feedback. For Head+D, participants mentioned that ``\textit{head shaking}" could lead to re-dwell (N=4) and that the technique could cause ``\textit{sore eyes}" or ``\textit{tiring eyes}" from prolonged use (N=2). As for Head+P, participants complained about ``\textit{poor gesture recognition}" (N=6) and the danger of becoming ``\textit{easily distracted}" (N=1).

Hand-based techniques were generally perceived to be ``\textit{novice and/or interesting}" (Hand+D: N=5; Hand+P: N=4), Hand+P was perceived to be ``\textit{convenient and easy to understand}" (N=3). However, like Head+P, participants complained about the gesture having ``poor recognition accuracy" (N=6) and being ``difficult to perform" (N=2). Hand+D was largely perceived to be ``\textit{tiring}" (N=6).

\section{Discussion}
\subsection{Performance}
Our results suggest that the four evaluated interaction techniques (Head+D, Head+P, Hand+D, Hand+P) have comparable task completion time. Head+D was found to be a better option than Hand+D regarding error rate. Prior work found that Dwell would outperform Pinch if using the same pointing technique \cite{mutasim2021pinch}; however, this is not supported by our findings. Our results suggest that Head+D did not outperform Head+P, nor did Hand+D outperform Hand+P. As expected, we observe that the Sentence Length could affect the task completion time, as participants needed to move further distances across the paragraphs when selecting longer text fragments. 

\subsection{User Experience and Workload}
We found Head+D to be the best technique among the 4 evaluated techniques. It outperformed both Hand+D and Hand+P on the average UEQ score and efficiency subscale, due to its good performance in task completion speed and accuracy. This was followed by Head+P, which outperformed Hand+D in the average UEQ rating. The poor ratings for Hand-based techniques could be due to hand/arm fatigue \cite{jang2017modeling} and tracking/recognition issues inherent in HoloLens 2 (supported by participants' comments), although it is one of the best AR HMDs in the market. Although we recommend Head+D among these 4 techniques, it is worth noting that none of these techniques, including Head+D, have any subscale reaching the ``Above Average" to ``Excellent" score compared to the benchmark scores provided by the UEQ tool \cite{UEQbenchmark}. This suggests that users felt the overall interaction was ``Bad" or ``Below Average" based on their prior experience with text selection on other platforms. Consequently, a better text selection technique is desirable.

We asked participants how comfortable they would feel about performing the techniques in public places and in front of others (e.g. co-workers, friends, and strangers). In general, participants said that they would be comfortable doing Head+D in front of others, as the technique is discreet. They said they would not mind involving Hand-based interaction in public places, but if they had a choice, they would choose not to. In terms of Selection Mechanism alone, participants said that they would be more comfortable using Dwell, especially with Head-based pointing. They stated that they would use Pinch when no other less conspicuous choices were available.  

Like user experience measurements, Head+D is generally the best technique for workload, having a lower mean workload and requiring less Temporal Demand and Effort than Hand+D. Head+D was also found to deliver better performance than Hand+P, which is in line with previous work \cite{6162910}.

\subsection{Usability Criteria (Revisited)}
We revisited the Usability Criteria scores based on our results. Instead of re-evaluating the selection mechanisms and pointing methods separately, we considered the four combined techniques that were evaluated in the user study. Of the two techniques using Dwell as their selection mechanism, Head+D received the highest ratings. For the most part, Dwell complemented Head-based pointing well, which resulted in the combined technique having the highest ratings across the four criteria. These findings were similar to the ratings for Dwell as a selection mechanism in standalone (see \autoref{table:CriteriaPre}). On the other hand, Hand+D did not attain the same positive results, as participants found that, in general, Hand-based interaction was not as usable---that is, the disadvantages of Hand-based pointing had a negative impact on the overall usability and performance of the combined technique. As shown in \autoref{table:Criteria}, both techniques that use pinch as their selection mechanism received a low rating for C1 and C4. However, Head+P received a low rating for C2, but not Hand+P. This is because users' hands would often move outside the active hand tracking area when used as a secondary input modality (i.e., for solely performing the gesture). This would occur as users tended to keep their hands in a more comfortable position \cite{arboundary,chiu2019pursuit}.

\begin{table}
\centering
  \caption{Ratings for each usability criterion for all techniques based on the results of our experiment (0-3 ticks indicate ratings from worst to the best). (C1) Simple, easy, and efficient to use; (C2) Minimal error rate and workload; (C3) Accessible and mobility enabling; and (C4) Socially acceptable.}
  \label{tab:freq}
  \begin{tabular}{p{1.5cm}p{1.2cm}p{1.2cm}p{1.2cm}p{1.2cm}}
    \toprule
    Mechanism& C1 & C2 & C3 & C4\\
    \midrule
    Head+Dwell & \checkmark \checkmark \checkmark & \checkmark \checkmark \checkmark & \checkmark \checkmark \checkmark & \checkmark \checkmark \checkmark \\

    Head+Pinch & \checkmark & \checkmark & \checkmark \checkmark \checkmark & \checkmark \\

    Hand+Dwell & - & - & \checkmark \checkmark \checkmark & \checkmark \\

   Hand+Pinch & \checkmark & \checkmark \checkmark & \checkmark \checkmark \checkmark & \checkmark \\

\bottomrule
\end{tabular}
\label{table:Criteria}
\end{table}

\subsection{Key Takeaways and Lessons}
\begin{itemize}
    \item \textit{Head+D} should be provided as the default interaction technique for text selection due to its relatively fast speed, lowest error rate, highest user experience, and lowest workload among the four evaluated techniques.
    
    \item \textit{Hand+P} is recommended as an alternative option, as opposed to Head+P, since participants mentioned that Hand+P was less distracting as their hands always remained within the hand tracking volume. 
    \item \textit{Head+P} and \textit{Hand+D} should only be used when no better options are available due to their poor ratings in all measures. 
    
\end{itemize}

\subsection{Limitations and Future Work}
One limitation is that we only considered distant interaction (i.e., out of users' arms' reach). However, there are cases where users need to interact within arm's reach (e.g., hold an instruction handbook). Future work will investigate suitable text selection techniques for different close-range text manipulation tasks. In addition, we did not consider eye gaze as a pointing mechanism as it was not that stable for fixations when reading small texts. As eye tracking and AR display technology advance further, we want to revisit this approach and explore it as part of our future work.

AR HMDs have been used as an external tool to support text manipulation tasks (i.e., reading, taking notes) \cite{holodoc}. In the future, building upon the findings obtained from this study, we plan to explore the integration of Head+D into the next steps of text manipulation, e.g., copy/cutting, deleting, and transferring selected texts to other platforms. As suggested by Microsoft, regarding the fundamentals of interaction in AR \footnote{https://docs.microsoft.com/en-us/windows/mixed-reality/design/interaction-fundamentals}, ``it's best to follow the guidance for a single model from beginning to end". 


\section{Conclusion}
This work has evaluated four text selection techniques, which were defined by combining different permutations of pointing methods (\textit{head}-based, \textit{hand}-based) and selection mechanisms (\textit{Dwell}, \textit{Pinch}), by conducting a study with 12 participants completing text selection tasks with a commercial Augmented Reality Head-Mounted Display. The experimental results show that \textit{Head}+\textit{Dwell} should be considered the first choice for text selection in such devices since it has a good performance (relatively fast speed and lowest error rate), provides the best user experience, is perceived to have the lowest workload, and is socially acceptable to use. Of the remaining techniques, \textit{Hand}+\textit{Pinch} can be provided as an alternative solution, while the results suggest the other two methods (\textit{Head}+\textit{Pinch}, \textit{Hand}+\textit{Dwell}) should be avoided. 


\begin{acks}
This work was supported in part by Xi'an Jiaotong-Liverpool University Key Special Fund (\#KSF-A-03), Natural Science Foundation of Guangdong Province (\#2021A1515012629), and Guangzhou Basic and Applied Basic Foundation (\#202102021131).
\end{acks}

\bibliographystyle{ACM-Reference-Format}
\bibliography{sample-base}




\end{document}